\documentclass[superscriptaddress,amsmath,amssymb,aps,prl,twocolumn,floatfix]{revtex4-2}

\usepackage{graphicx}
\usepackage{float}
\usepackage{bm}
\usepackage[colorlinks, linkcolor=mycol, citecolor=mycol, urlcolor=mycol, breaklinks]{hyperref}
\usepackage{braket}
\usepackage{bbold}
\usepackage{xcolor} 
\usepackage{comment}
\usepackage{physics}
\usepackage{verbatim}

\newcommand{\mi}{ {\rm i} }
\newcommand{\me}{ {\rm e} }
\newcommand{\id}{\mathbb{1}}

\usepackage{amsmath}
\usepackage{amssymb}
\usepackage{ulem}

\definecolor{mycol}{RGB}{10,55,130}

\definecolor{Myblue}{HTML}{648fff}
\definecolor{Myred}{HTML}{cc6677}
\definecolor{Mygreen}{HTML}{558B2F}
\newcommand{\customsection}[1]{\textit{#1---}}

\begin{document}

\title{Unraveling Dicke Superradiant Decay with Separable Coherent Spin States} 

\author{P.~Rosario}
\affiliation{Departamento de Física, Universidade Federal de S\~ao Carlos, Rodovia Washington Lu\'is, km 235—SP-310, 13565-905 S\~ao Carlos, SP, Brazil}
\affiliation{CESQ/ISIS (UMR 7006), CNRS and Universit\'{e} de Strasbourg, 67000 Strasbourg, France}

\author{L.~O.~R.~Solak}
\affiliation{Departamento de Física, Universidade Federal de S\~ao Carlos, Rodovia Washington Lu\'is, km 235—SP-310, 13565-905 S\~ao Carlos, SP, Brazil}
\affiliation{CESQ/ISIS (UMR 7006), CNRS and Universit\'{e} de Strasbourg, 67000 Strasbourg, France}

\author{A.~Cidrim}
\affiliation{Departamento de Física, Universidade Federal de S\~ao Carlos, Rodovia Washington Lu\'is, km 235—SP-310, 13565-905 S\~ao Carlos, SP, Brazil}

\author{R.~Bachelard}
\affiliation{Departamento de Física, Universidade Federal de S\~ao Carlos, Rodovia Washington Lu\'is, km 235—SP-310, 13565-905 S\~ao Carlos, SP, Brazil}

\author{J.~Schachenmayer}
\affiliation{CESQ/ISIS (UMR 7006), CNRS and Universit\'{e} de Strasbourg, 67000 Strasbourg, France}
\thanks{schachenmayer@unistra.fr}

\date{\today}

\begin{abstract}
We show that idealized Dicke superradiant decay from a fully inverted state can at all times be described by a positive statistical mixture of coherent spin states (CSS). Since CSS are separable, this implies that no entanglement is involved in Dicke decay. Based on this result, we introduce a new numerical quantum trajectory approach leading to low-entanglement unravelings. This opens up new possibilities for employing matrix product state (MPS) techniques for large-scale numerical simulations with collective decay processes.
\end{abstract}

\maketitle

\customsection{Introduction}The Dicke model \cite{dicke1954coherence}, a cornerstone of quantum optics, describes the dynamics of a large, all-to-all-coupled ensemble of $N$ two-level atoms resulting from their coupling to the same single mode of the electromagnetic field. The model predicts a phenomenon coined as pure superradiance: Atoms spontaneously emit light cooperatively, leading to an intense, short burst of radiation whose intensity scales with $N^2$~\cite{gross1982superradiance}. Although semiclassical approaches have successfully described aspects of superradiance (e.g., burst times $t_B \propto \ln N$)~\cite{MacGillivray1976,Arecchi1972}, hinting at the classicality of the phenomenon, understanding whether quantum correlations are generated throughout Dicke dynamics has long remained a challenge~\cite{2018_Tura_Sanpera_quantum, 2009_Toth_Guhne_apb, 2014_Wolfe_Yelin_prl,Nengkun_separability_2016, amico2008entanglement}. Today, a common picture is to consider the many-body state evolving in time by following a cascade over states symmetric under atom exchange. Although these Dicke states are individually entangled, a general solution for the dynamics is a mixture of them. Currently, extensive works have identified explicit new solutions of the dynamics~\cite{holzinger2025solving,holzinger2024exact,2022_Malz_Cirac_pra}, yet asserting their separability is considered NP-hard~\cite{2018_Tura_Sanpera_quantum,2009_Toth_Guhne_apb}. Thus, given the inherited complexity of the task, it has been conjectured, and numerically shown only for a few emitters, that no entanglement should be generated~\cite{2014_Wolfe_Yelin_prl,Nengkun_separability_2016}.

\smallskip

In this work, we introduce a decomposition of the time-dependent state into a mixture of coherent spin states (CSS). CSS are product states, implying that, as long as the mixture is positive, no entanglement is present. We develop a numerically efficient technique to find positive mixtures --- which we test up to $N=50$. This insight is used to devise
a new entanglement-optimized quantum trajectory algorithm that leads to unravelings of dynamics into almost perfect CSS trajectories. These results pave the way for efficient large-scale simulations of open systems with collective processes.

\begin{figure}
    \centering
    \includegraphics[width=0.49\textwidth]{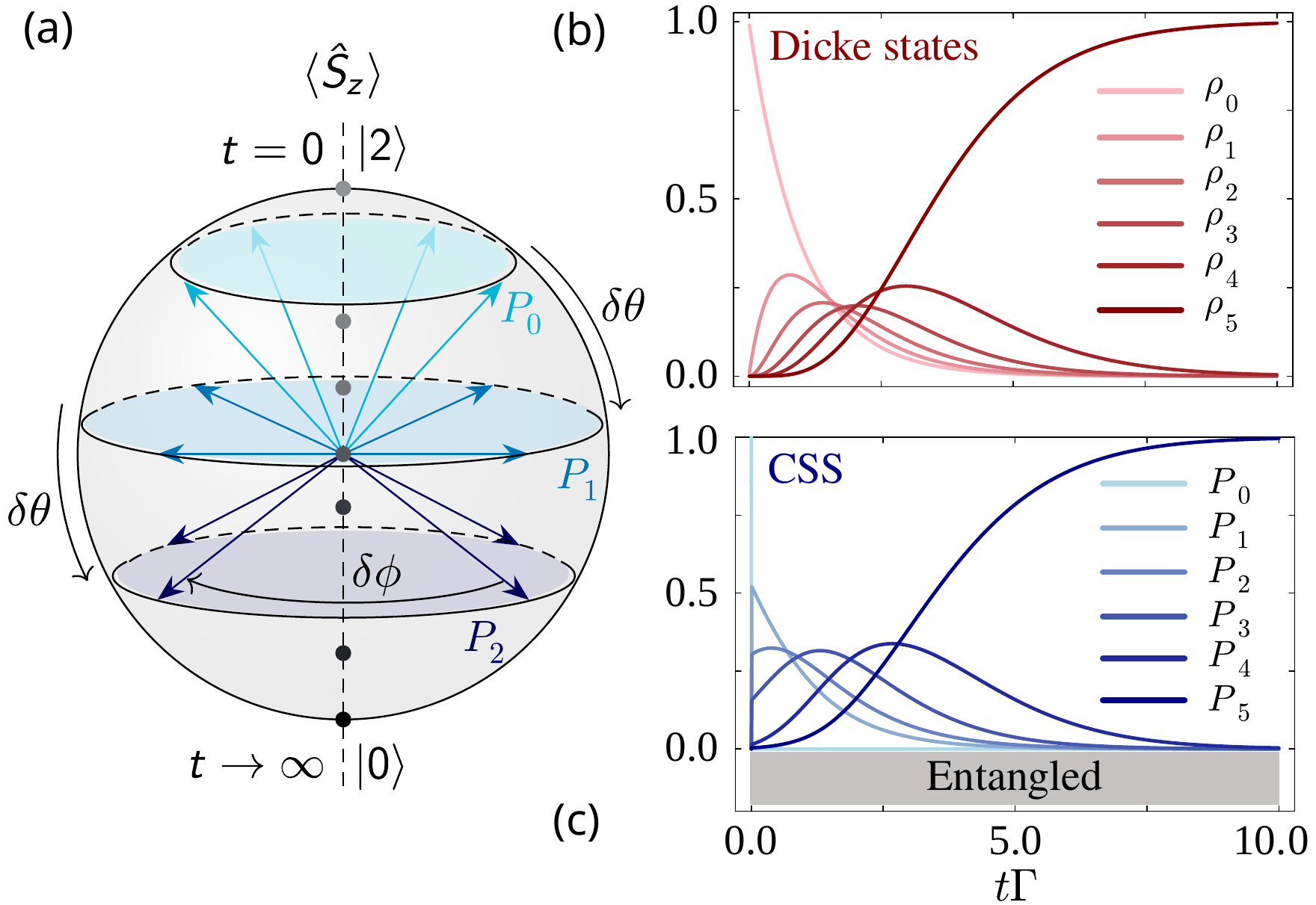}
    \caption{(a) Superradiant decay represented as a fall through the collective Bloch sphere. This leads to mixed states (grey points) on the $S_z$-axis. Our ansatz consists in representing those states as mixtures of coherent spin states (CSS), Eq.~\eqref{eq:CSS_ansatz}, here sketched for $N=2$ (12 arrows pointing to the surface of the sphere). (b) Evolution of the coefficients $\rho_m$ in the entangled Dicke state expansion~\eqref{eq:DS_ansatz} ($N=5$). (c) Equivalent positive evolution of expansion coefficients $P_a$ in the CSS ansatz~\eqref{eq:CSS_ansatz}, which demonstrates the absence of entanglement.}
    \label{fig:intro}
\end{figure}

\smallskip

\customsection{Superradiant Dicke cascade}In the idealized Dicke problem, $N$ two-level emitters collectively decay from an excited state $\ket{e}$ to a ground state $\ket{g}$. The decay dynamics is described by the Lindblad master equation
\begin{align}
    \label{eq:dicke_meq}
    \frac{d}{dt} \hat \rho = 
    {\Gamma} 
    \left[
    2  \hat L \hat \rho \hat L^\dag 
    -  \hat \rho \hat L^\dag \hat L
    -  \hat L^\dag \hat L \hat \rho 
    \right],
\end{align}
with the collective jump operator $\hat L =  \sum_i \hat \sigma_i /\sqrt{2N}$, $\hat \sigma_i = \dyad{g}{e}_i$ the single-emitter lowering operators, and $\Gamma$ the collectively enhanced decay rate. We consider a fully inverted initial state $\hat \rho_0 = \bigotimes_{i=1}^{N}\dyad{e}{e}_i$~\cite{1977_Lee_pra}. In our idealized setup of Eq.~\eqref{eq:dicke_meq}, no Hamiltonian is present. Instead, all emitters couple identically to the same mode. This model was originally envisioned as arising from a densely packed set of two-level emitters with distances very small compared to the wavelength of the transition, neglecting coherent Hamiltonian contributions~\cite{lehmberg1970radiation,gross1982superradiance}. In more practical setups, such an idealized model can be realized using atoms in optical cavity systems \cite{Thompson_2016,Majer_2018}, near 1D waveguides \cite{Philipp_2024}, or even with atomic nuclei excited by free-electron lasers \cite{Tetsuya_2018}, to name just a few examples.


Permutation symmetry allows for numerical simulations of very large systems~\cite{Shammah2018} and enables analytical insight~\cite{holzinger2024exact, holzinger2025solving}.  A common way to solve the dynamics is to restrict evolution to the symmetric Dicke basis $\ket{m}$,
\begin{align}
    \hat \rho(t) = \sum_{m=0}^N \rho_m (t) \dyad{m}{m}.
    \label{eq:DS_ansatz}
\end{align}
For $N$ emitters, $m=0,\dots,N$ is the number of emitters in the excited state $\ket{e}$, and $\ket{m}$ is the symmetrized wave-function of all such states ~\cite{2014_Wolfe_Yelin_prl, holzinger2024exact}. Dicke states are collective spin states of a spin-$S$ model with $S=N/2$, and the spin-lowering operator acts on the Dicke basis as $\hat S^-\ket{m} = \sqrt{S(S+1) - (S - m) (S + 1 - m)} \ket{m-1}$. In our convention, $\hat L = \hat S^- / \sqrt{2N}$. Hence, the dynamics can be visualized in a collective Bloch sphere, where the decay from the fully inverted state evolves through the center of the sphere [Fig.~\ref{fig:intro} (a)]. An evolution of Dicke state probabilities $\rho_m(t)$ is shown in  Fig.~\ref{fig:intro}(b) for $N=5$.

\smallskip
\customsection{Entanglement}Dicke states with $m \in [1, N-1]$ are entangled~\cite{latorre2005entanglement, schachenmayer2013entanglement, toth2007detection, duan2011entanglement, lucke2014detecting}, and for any bipartition of emitters into two blocks $A$ and $B$, they feature a finite entanglement entropy. The von Neumann entanglement entropy is defined as $S_{\rm VN}(\ket{\psi}) = - \sum_\beta^{D_B} s_\beta \log_2(s_\beta)$ for a pure state $\ket{\psi}$. Here, $s_\beta$ are the eigenvalues of the reduced density matrix in block $B$, obtained from the partial trace $\hat \rho_B = {\rm tr}_A(\dyad{\psi}{\psi})$, and $D_B$ is the Hilbert space dimension of block $B$. For a single Dicke state and a splitting of emitters into two blocks with $N_B$ (block $B$) and $N-N_B$ (block $A$), one finds $ s_{\beta} = { {N_B \choose \beta} {N-N_B \choose m-\beta}}/{{N \choose m}}$ and $D_B = N_B+1$~\cite{latorre2005entanglement, SM}. For a symmetric state $S_{\rm VN}$ is the largest for $N_B = N/2$, we thus focus below on this splitting. The entanglement in the half-excited Dicke state is proportional to $S_{\rm VN}(\ket{m=N/2}) \propto \log(N)$~\cite{latorre2005entanglement, schachenmayer2013entanglement}, growing slowly with $N$ due to the symmetry-restricted sub-Hilbert space size $D_B = N/2 + 1 \ll 2^N$ for large $N$.

\smallskip

\begin{figure}[t]
    \centering
    \includegraphics[width=1\linewidth]{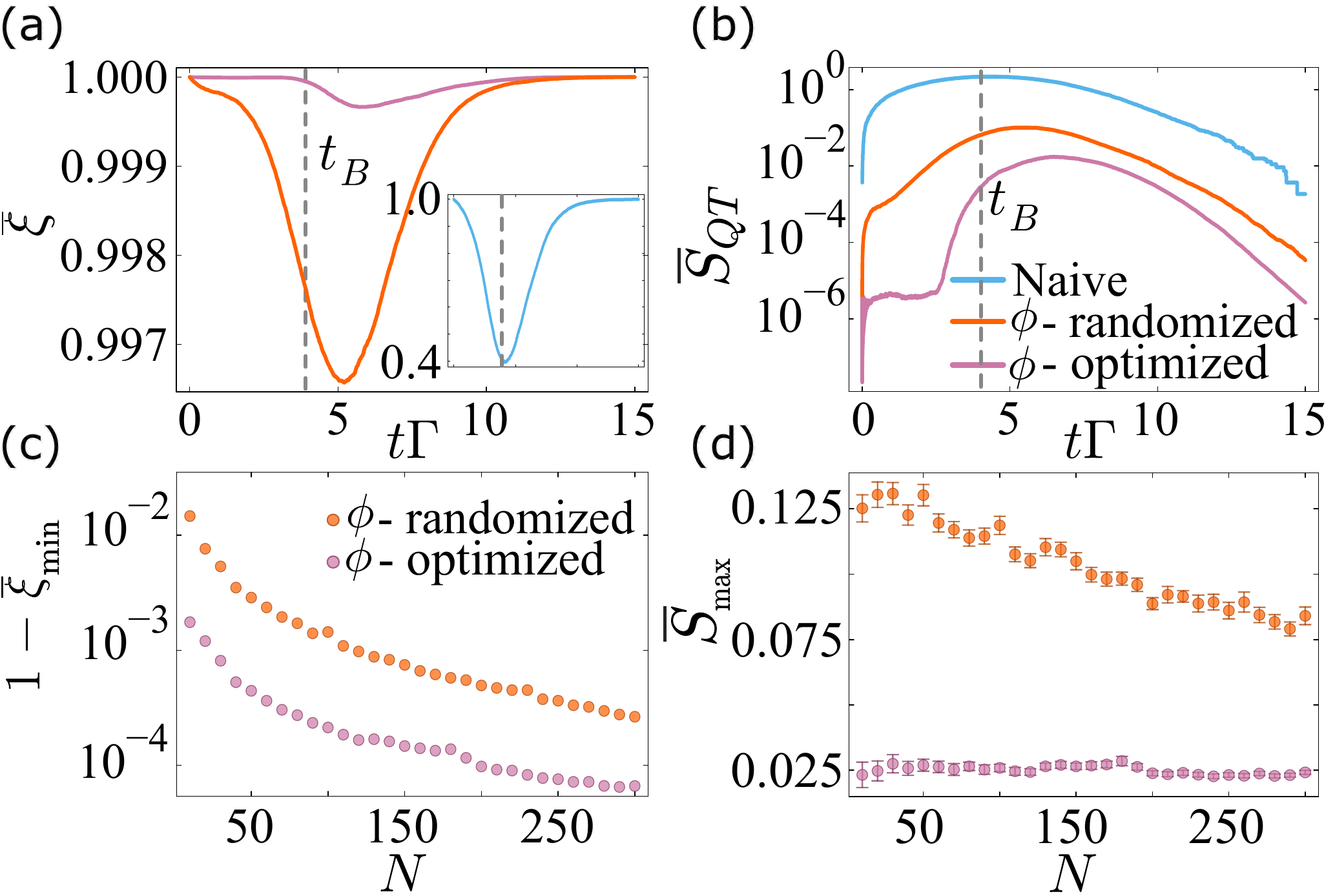}
    \caption{Trajectory averaged evolution ($N=50$) of (a) normalized Bloch vector length $\overline{\xi}$ and (b) TE. We compare naive [Kraus operators \eqref{eq:kraus_E}], $\phi$-randomized, and $\phi$-optimized unravelings [using randomized/optimized choices of Kraus operators~\eqref{eq:kraus_F}, see text]. The scaling with $N$ is analyzed for averaged: (c) minimum Bloch vector length, $\xi_\chi$; and (d) maximum TE, $\overline{S}_{\rm max}$. Dashed lines highlight the burst time.}
    \label{fig:qt}
\end{figure}

For a statistical mixture of pure states $|\psi^{[\chi]} \rangle$, $\hat \rho = \sum_{\chi} q_\chi |\psi^{[\chi]} \rangle \langle\psi^{[\chi]}|$ with $q_\chi > 0$ and $\sum_\chi q_\chi = 1$, we define an averaged trajectory entanglement (TE) as $\overline{S}_{\rm QT}(\hat \rho) = \sum_{\chi} q_\chi S_{\rm VN}(|\psi^{[\chi]} \rangle)$. For example, when computing superradiant decay in the Dicke expansion [Eq.~\eqref{eq:DS_ansatz}] one finds that TE is finite, $\overline{S}_{\rm QT}(\hat \rho(t)) = \sum_m \rho_m (t) S_{\rm VN}(\ket{m}) > 0$, except for $t=0$ and $t \to \infty$, see Fig.~\ref{fig:qt}. Nevertheless, TE is \textit{not} connected to genuine separability~\cite{daraban2025non, preisser2023comparing} since the decomposition into pure states is highly non-unique. A proper definition of entanglement is given by the entanglement of formation~\cite{bennett_mixed-state_1996}, which is defined as the minimum TE of all pure-state ensembles representing the density matrix. Finding such a minimium is generally considered NP-hard~\cite{gurvits2003,gharibian2010}. We will now introduce an alternative expansion into well-chosen CSS with vanishing TE, thus zero entanglement of formation at all times.

\smallskip

\customsection{Separable CSS ansatz}Let us consider the expansion
\begin{align}
    \hat \rho(t) = \sum_{a=0}^{N} P_a(t) 
    \left[\frac{1}{N_\phi}\sum_{b=1}^{N_\phi}
    \dyad{\theta_a(t), \phi_b}{\theta_a(t), \phi_b}
    \right],
    \label{eq:CSS_ansatz}
\end{align}
where $\ket{\theta_a(t),\phi_b}$ denotes a CSS parametrized by a (time-varying) polar angle $\theta_a(t)$ and an azimuthal angle $\phi_b$. A CSS is a product state, $\ket{\theta_a(t),\phi_b}= \bigotimes_j [\cos{\theta_a(t)} \ket{g}_j + \exp(\mi \phi_{b}) \sin(\theta_a)  \ket{e}_j]$~\cite{Arecchi1972,A_M_Perelomov_1977}, represented by points on the Bloch sphere surface. In Eq.~\eqref{eq:CSS_ansatz} the angles $\phi_b$ are chosen uniformly such that on average the $S_x$ and $S_y$ components vanish [see Fig.~\ref{fig:intro}(a)]. We choose $N_\phi = 2N$ angles $\phi_b = b\,(\pi/N) = b\,\delta \phi$ ($b=0,\dots,N_\phi$), but note that larger $N_\phi = x N$ with integer $x > 2$ are also possible. The vector $\bm{P}_t$ with elements $P_a(t)$ represents the time-dependent state as a distribution of polar angles. Note that our ansatz is reminiscent of a CSS phase-space description~\cite{Narducci1974}. There, states in the Dicke cascade are described by a time-dependent Wigner function, similar to $P_a(t)$. A Wigner function, however, can become negative, while here we aim at finding $P_a(t)\geq 0$.

\begin{figure*}
    \centering
    \includegraphics[width=\textwidth]{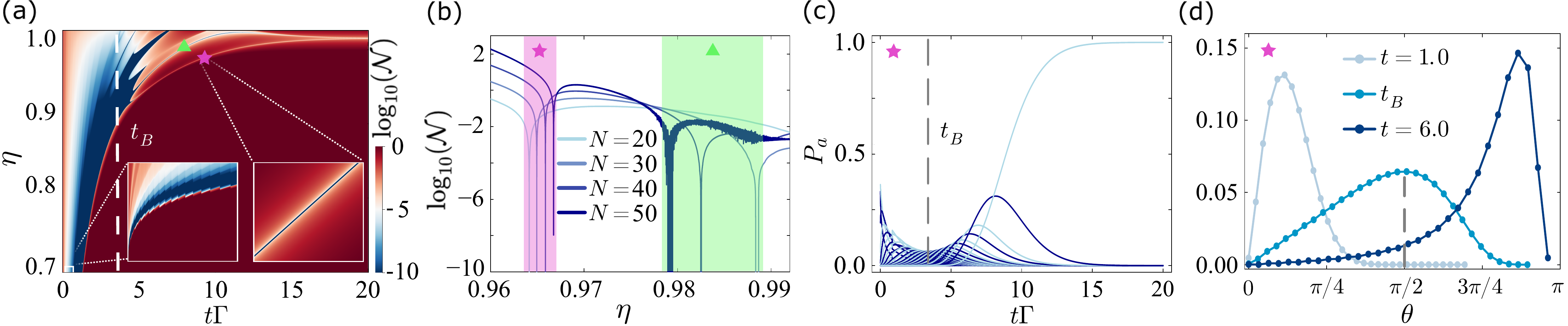}
    \caption{(a)  Logarithm of negativity $\log_{10}(\mathcal{N})$ in the range $[-10,0]$ as function of time and $\eta$, for $N=30$. Before the burst time $t_B = \ln (N)/\Gamma$ (dashed lines) many positive solutions exist. Insets: Zooms in the ranges $0 \leq t\Gamma \leq 0.5$, $10^{-3} \leq \eta \leq 0.7$ and $9 \leq t\Gamma \leq 9.02$, $0.97332 \leq \eta \leq  0.97344$. After $t_B$, two passages to late times with positive solutions are highlighted: star (lower, at all time remaining $\eta \leq 1$), triangle (upper). (b) $\log_{10}(\mathcal{N})$ as function of $\eta$ at $t\Gamma=8$. For different $N$, $\mathcal{N} \to 0$ in the areas of the two passages. (c)  Evolution with $P_a >0$ for a numerically extracted $\eta(t)$ along the lower passage ($N=30, \mathcal{N} < 10^{-6}$). (d)  $\{P_{a}\}_{a=0}^{N}$ distribution as function of $\theta$ at different times ($N=30$). At $t_B$, the peak centers around $\theta=\pi/2$.}
    \label{fig:eta_plot}
\end{figure*}

\smallskip

\customsection{Positive solutions with the CSS ansatz}In the Dicke state expansion~\eqref{eq:DS_ansatz}, a solution to Eq.~\eqref{eq:dicke_meq}  at time $t$ is described by a $(N+1)$-dimensional vector $\bm{\rho}(t)$ with elements $\rho_m(t)$. It is given by $\bm{\rho}_t = \bm{A}_t[:,1]$, where $\bm{A}_t[:,1]$ denotes the first column of the evolution matrix $\bm A_t=\exp({\bm D t})$. The $(N+1) \times (N+1)$ matrix $\bm{D}$ has a bidiagonal form~\cite{holzinger2025solving},
$(\bm{D})_{mn} = -\delta_{mn} \beta^{2}_{N-m}  + \delta_{m(n-1)} \beta^{2}_{N-m},$ for $m,n=0,\dots,N$, with $\beta_m^2 = m (N-m + 1)/N$  and $\delta_{mn}$ the Kronecker delta. Note that $\bm{A}_t$ is a stochastic matrix in the sense of Markov chains, satisfying the conditions $0\leq (\bm{A}_t)_{mn}\leq 1$ and $\sum_{m=0}^{N}(\bm{A}_t)_{mn}=1, \forall n$. A key feature of a mapping with a stochastic matrix is that $\bm{\rho}_t$ remains positive and normalized, $\sum_m \rho_m(t)=1$. Thus, $\bm{\rho}_t$ remains a valid probability distribution during the dynamics.

\smallskip

We introduce a mapping between the representations,
\begin{align}
    \bm{\rho}_t = \bm{M}_t \cdot \bm{P}_t \ \quad \ \bm{P}_t = \left(\bm{M}_t\right)^{-1} \cdot  \bm{\rho}_t,
    \\ (\bm{M}_t)_{mn} =\begin{pmatrix} N\\ m  \end{pmatrix}z_n(t)^{N-m}[1-z_{n}(t)]^{m}.
\end{align}
Here, we define a time-dependent mapping matrix $\bm{M}_t$ with $N+1$ time-dependent functions $z_a(t) = \cos^2(\theta_a(t)/2) \in [0,1]$ \cite{SM}. Note that $\bm{M}_t$ is a stochastic matrix, guaranteeing that $\bm{\rho}_t$ is positive if $\bm{P}_t >0$. However, the inverse matrix $(\bm{M}_t)^{-1}$ is not necessarily stochastic. To prove that $\bm{P}_t$ remains positive, one can show that the $\bm{P}$-evolution matrix, $(\bm{M}_t)^{-1}  \cdot  \bm{A}_t$, is stochastic. It is possible to achieve this iteratively: Defining a discrete time index $\tau$, such that $t=\tau \Delta t$, and $\bm{A}^{[\Delta t]} = \exp(\bm{D} \Delta t)$, such that $\bm{A}_t = (\bm{A}^{[\Delta t]})^\tau$, in first order it can be shown that $ (\bm{M}_t)^{-1} \cdot \bm{A}^{[\Delta t]} \cdot \bm{M}_t$ preserves the positivity of $\bm{P}$ for $\Delta t \to 0$ (see companion article~\cite{bassler2025absence}).

\smallskip

Let us now construct an explicit positive solution for $\bm{P}_t$, which requires solving a system of linear equations
\begin{align}
    \bm{M}_t \cdot \bm{P}_t = \bm{A}_t[:,1].
    \label{eq:linsys}
 \end{align}
Our objective is to solve this linear system with a positive vector $\bm{P}_t$. Then, since $\bm{A}_t[:,1]$ is a solution to Eq.~\eqref{eq:dicke_meq} in the Dicke state representation at time $t$, we have shown that there exists a positive solution to the Dicke dynamics also in the CSS basis~\eqref{eq:CSS_ansatz}. The transformation matrix $\bm{M}_t$ depends on $N+1$ parameters, finding a positive $\bm{P}_t$ is thus a multi-parameter optimization problem. We now introduce a key step that can turn it into a simpler single-parameter problem: We restrict $\bm{M}_t$ to matrices with equally spaced polar angles $\theta_{a}(t)=\eta(t) \, a\pi/N$, using a single parameter $\eta(t)$. $\bm{M}_t$ then belongs to the family of centrosymmetric matrices. For such matrices the inverses always exist~\cite{1970_I_J_Good}, and $\eta(t)$ can in principle be an arbitrary positive number, corresponding to a set of CSS angles $\theta$ (interpreted in the range $[0, \pi]$). The challenge is to find an $\eta(t)$ that leads to positive $\bm{P}$ at all times. 

Note that $\eta(t)$ does not need to be a smooth function, and multiple positive-$\bm{P}$ solutions can exist. However, for $N=2$, we identify the following analytical one~\cite{SM}:
\begin{align}
\label{eq:eta_n2analytic}
\eta(t)=\frac{2}{\pi}\arccos{\sqrt{\frac{t\Gamma\me^{-t\Gamma}}{2(1-\me^{-t\Gamma})-t\Gamma \me^{-t\Gamma}}}}.
\end{align}
Obtaining $\eta(t)$ analytically for larger $N$ requires solving a set of transcendental inequalities~\cite{SM}.

To find positive $\bm{P}_t$ numerically, we introduce the negativity $\mathcal{N} = - \sum_a \min(P_a, 0)$. In Fig.~\ref{fig:eta_plot}(a) we plot $\log_{10}(\mathcal{N})$ as a function of time $t$ and $\eta$ for $N=30$. Note that we cut off values $\mathcal{N} < 10^{-10}$ and $\mathcal{N}>1$ in order to avoid showing effects of machine precision noise and to ignore regions with diverging negativity. Below the burst time, $t_B = \ln(N)/\gamma$,  $0 < t \leq  t_B $ we find a large parameter range (blue area) with many positive solutions, extending to $t \to 0$ (left inset). Note that at $t=0$, $P_a = \delta_{a0}$ is a trivial solution. For $t>t_B$, the number of solutions reduces, but we still find several positive ``passages'' to late times. We highlight two passages with a star (lower) and a triangle (upper). In general, passages are very narrow, but zooming into the respective regions (right inset) reveals negativities that can be made arbitrarily small. This point can be better appreciated in Fig.~\ref{fig:eta_plot}(b), where a slice of $\mathcal{N}$ is shown, as a function of $\eta$ at time $t\Gamma = 8$. We observe $\mathcal{N} \to 0$ at the locations of the passages (up to $N=50$, where noisy behavior due to machine precision limitations starts to appear). Although the upper passage has some internal structure, it appears that the lower passage corresponds to a single solution with $\eta(t) \leq 1$ at all times~\cite{SM}. Note that for $N=2$ the lower passage corresponds exactly to Eq.~\eqref{eq:eta_n2analytic}. Due to the narrow nature of the passage, it is difficult to find a smooth fitting function with low $\mathcal{N}$. We can however extract solutions by minimizing $\mathcal{N}$ numerically along the lower passage~\cite{SM}. We note that finding a smooth solution of $\eta(t)$ numerically that fully follows the lower passage becomes more difficult with $N$ due to machine precision limitations in the linear solver. However, solutions that are allowed to jump between the passages are generally much easier to find. In addition, we have observed the existence of positive solutions for values of $\eta(t) > 1$. For a solution along the lower passage, in Fig.~\ref{fig:eta_plot}(c) we show a corresponding positive evolution of $P_a$ as a function of time (for $N=30$). We always observe that at $t_B$, $\max(\bm{P}_{t_B})$ is lowest in the evolution~\cite{SM}. In Fig.~\ref{fig:eta_plot}(d) we plot the distribution of $P_a$ at various $t$, showing that $\max(\bm{P}_{t_B})$ corresponds to $\theta = \pi/2$. This asserts that CSS populations are distributed around the equator of the Bloch sphere at the time of the burst. Having established that a positive CSS decomposition exists, we now introduce an algorithm finding trajectory unravelings close to CSS.

\smallskip

\customsection{Quantum trajectory unraveling}Our quantum trajectory (QT) method emerges from a Kraus operator formalism~\cite{daraban2025non, vovk2024quantum}. Markovian dynamics allows for a discretized evolution with timesteps $\Delta t \to 0$. Each step can be formulated as a quantum operation  $\mathcal{E}(\bullet)$  with  operator sum representation $\hat\rho_{\rm out} = \mathcal{E}(\hat \rho_{\rm in}) = \sum_k \hat E_k \hat \rho_{\rm in} \hat E_k^\dag$~\cite{nielsen2010quantum}. For Eq.~\eqref{eq:dicke_meq} there are two Kraus operators
\begin{align}
\label{eq:kraus_E}
    \hat E_0 = \id - \Delta t \Gamma \hat L^\dag \hat L, \text{ \ \quad  \ }
    \hat E_1 = \sqrt{2 \Delta t \Gamma} \hat L.
\end{align}

In QT algorithms, the time-dependent density matrix $\hat \rho_t$ is approximated by a large number $n_T$ of pure-state samples, $\hat \rho_t \approx \sum_{\chi = 1}^{n_T} |\psi^{[\chi]}_t \rangle \langle\psi^{[\chi]}_t| / n_T$, with the initial condition $|\psi^{[\chi]}_0\rangle = |{m=N}\rangle$ for all $\chi = 1,\dots,n_T$. A single timestep, which evolves a trajectory from $|\psi^{[\chi]}_t\rangle$ to $|\psi^{[\chi]}_{t+\Delta t}\rangle$, is stochastically simulated by applying $\hat E_k$ with probability $p_k = \langle{\psi^{[\chi]}_t}| \hat E_k^\dag \hat E_k |{\psi^{[\chi]}_t}\rangle$, followed by a renormalization, $|{\psi^{[\chi]}_{t+\Delta t}}\rangle = \hat E_k |{\psi^{[\chi]}_{t}}\rangle/\sqrt{p_k}$. As a result, the statistical approximation of $\hat \rho_{t+\Delta t}$ is exact for $n_T \to \infty$, since $\mathcal{E}\left(|{\psi^{[\chi]}_{t}}\rangle \langle{\psi^{[\chi]}_{t}}|\right) = \sum_k p_k \hat \rho_k$ with $\hat \rho_k = \hat E_k |{\psi^{[\chi]}_{t}}\rangle \langle{\psi^{[\chi]}_{t}}| \hat E_k^\dag / \langle{\psi^{[\chi]}_{t}}| \hat E_k^\dag\hat E_k |{\psi^{[\chi]}_{t}}\rangle$ and $p_k$ a probability distribution. A quantum operation acting on $|{\psi^{[\chi]}_{t}}\rangle \langle{\psi^{[\chi]}_{t}}|$ thus amounts to randomly replacing the state $|{\psi^{[\chi]}_{t}}\rangle \langle{\psi^{[\chi]}_{t}}|$ with $\hat \rho_k$. In the limit $\Delta t \to 0$, this approach is equivalent to the standard QT method developed in quantum optics~\cite{dalibard_wave-function_1992, dum_monte_1992, carmichael1993an, gardiner_wave-function_1992}, which has been key to efficient simulations of open quantum many-body systems~\cite{daley2014quantum}. 

\smallskip

Naively performing QT with Kraus operators from Eq.~\eqref{eq:kraus_E} restricts all trajectories to the Dicke basis and thus to the $S_z$-axis of the Bloch sphere. The inset in Fig.~\ref{fig:qt}(a) confirms this, showing an evolution of the normalized Bloch vector length, $\xi_\chi = 2{\langle{\psi^{[\chi]}_{t}} |\hat{\mathbf{S}}^2|{\psi^{[\chi]}_{t}} \rangle} ^{1/2}/N$, from $ 1 \to 0 \to 1$ (we denote a trajectory average by $\overline{\bullet}$). Using the naive unraveling also implies that trajectories become entangled. Around $t_B$, with most of the population in states $\ket{m\approx N/2}$, TE is thus of the order $\overline{S}_{\rm QT} \sim \log N$~\cite{latorre2005entanglement, schachenmayer2013entanglement}, as confirmed in Fig.~\ref{fig:qt}(b). From a classical computational point of view, this is inefficient, as above we have shown that TE can be zero. Let us now present a method to drastically reduce TE.

\smallskip

\customsection{Low-entanglement QT unraveling}We exploit  the unitary degree of freedom of Kraus operators~\cite{vovk2024quantum, daraban2025non},
\begin{align}
    \label{eq:kraus_F}
    \hat F_n = \sum_k u_{nk} \hat E_k.
\end{align}
Here, $\hat F_n$ ($n,k=0,1$) are Kraus operators defining the same quantum operation as $\hat E_k$, as long as the matrix with elements $(\bm{u})_{nk} = u_{nk}$ is unitary~\cite{nielsen2010quantum}. Any observable evolution can be computed in QT with Kraus operators $\hat F_n$ instead of $\hat E_k$. Entanglement entropy, however, is not an observable but a nonlinear function of the state. As such, it can substantially depend on the choice of $\bm{u}$~\cite{vovk2024quantum, daraban2025non}. Following~\cite{daraban2025non} we focus on unitary matrices parametrized with $\theta_F \in [0, \pi/2]$ and $\phi_F \in [0, 2\pi]$:
\begin{align}
  \bm{u}\left( \theta_F,\phi_F \right)= \begin{pmatrix} \cos\theta_F& \sin\theta_F\\ -\sin\theta_F& \cos\theta_F \end{pmatrix} \begin{pmatrix} \me^{\mi\phi_F}& 0\\ 0& \me^{-i\phi_F} \end{pmatrix}.\label{eq:u_param}
\end{align}
In the limit of $N \to \infty$, this parametrization leads to
\begin{align}
    \label{eq:F_largeN}
    &\hat{F}_{n}\to \me^{(-1)^n\exp({-\mi2\phi_{F}})\cot(\theta_{F})^{2n-1}\sqrt{2\Delta t}\hat{L}}.
\end{align}   
A Bloch vector rotation can be performed with $\hat R(\theta, \phi) = \me^{\exp(\mi\phi)\tan(\theta/2)\hat S^-}$~\cite{A_M_Perelomov_1977}. Therefore, we observe that for large $N$, applying a Kraus operator $\hat F_n$ to a CSS  (without renormalization) becomes equivalent to mapping it to another CSS. Note that the polar rotation induced by the Kraus operator depends on $\sqrt{\Delta t}$ and $\theta_F$. We find that the precise choice of $\theta_F$ does not have any major impact on TE \cite{SM} (except close to the trivial $\theta_F=0,\pi$). We therefore fix $\theta_F=\pi/4$ while varying $\phi_F$ only. Note that for $N \to \infty$, $\phi_F \to \phi/2$. In ansatz~\eqref{eq:CSS_ansatz}, the  $\phi_b$ are homogeneously distributed. Therefore, to lower TE we propose to randomly select $\phi_F$ uniformly at each quantum operation. In Fig.~\ref{fig:qt} we demonstrate that this \textit{$\phi$-randomized} algorithm leads to a strong reduction of TE. In addition, the trajectory Bloch vectors remain now almost on the sphere as $\overline{\xi} \to 1$. With increasing $N$, spin-lengths further increase while TE decreases, as shown in Fig.~\ref{fig:qt}(c/d). There we plot the averaged maximum (minimum)  von Neumann entanglement entropy $\overline{S}_{\rm max}$ (spin length $\overline{\xi}_{\rm max}$), taken over time, [$S_{\rm max} = \max_t(S_{\rm VN}(|\psi^{[\chi]}_{t}\rangle)$, 
$\xi_{\rm min} = \min_t(\xi_\chi)$]. This is in line with the expectation from Eq.~\eqref{eq:F_largeN}.

\smallskip

We can further reduce TE by resorting to an adaptive minimization~\cite{vovk2024quantum, daraban2025non}. Defining $|\psi^{[\chi]}_{\rm po} \rangle_{0,1} = \hat F_{0,1} |{\psi^{[\chi]}_t}\rangle$, the post-operation entanglement is $S_{\rm po}(\phi_F) = [ p_0 S_{\rm VN}(|\psi^{[\chi]}_{\rm po} \rangle_0) + p_1 S_{\rm VN}(|\psi^{[\chi]}_{\rm po} \rangle_1]/2$. Using expressions from~\cite{SM}, we use $S_{\rm po}(\phi_F)$ as a cost function for numerical minimization at each step. After finding $\phi_F^o$ minimizing $S_{\rm po}(\phi_F)$, we use a QT step with the corresponding $\hat F_{0,1}(\phi_F^o)$. We find that this \textit{$\phi$-optimized} algorithm leads to TE below $10^{-5}$ for $N=50$ before $t_B$, and always two orders of magnitude smaller than the naive unraveling [Fig.~\ref{fig:qt}(b)]. Now, $1-\overline{\xi}_{\rm min} \lesssim 10^{-3}$ for $N=50$ and $\overline{\xi}_{\rm min} \to 1$ with increasing $N$, indicating an a convergence towards CSS trajectories for large $N$. In Fig.~\ref{fig:qt}(d) we observe no $N$ dependence on the residual $S_{\rm max}$. Note that for $N>300$ also in the $\phi$-randomized algorithm the scaling with $N$ becomes sub-linear, e.g.~for $N=800$ we find $\overline{S}_{\rm max}=0.0612\pm 0.003$ (not plotted). We attribute this to the renormalization step in QT. We also note that the $\phi$-optimization is not guaranteed to find zero TE due to the ``greedy'' nature of the algorithm, only considering knowledge of a single trajectory at a time. Nevertheless, it is important to highlight that such entanglement entropies are extremely small. In practical scenarios, these low entropies will allow us to describe states almost as products, enabling the use of MPS with extremely small bond dimensions $\chi$~\cite{schuch2008entropy, schollwock2011density}. While the bond dimensions usually depend exponentially on the entanglement entropies ($\chi \sim 2^{\overline{S}_{\rm max}}$), here $\chi \approx 2$ would already provide a close-to-perfect state approximation. Our unraveling scheme will therefore grant the removal of nearly all trajectory entanglement contributions from symmetric Dicke decay process, and can thus drastically improve the efficiency of simulations of quantum dynamics in more general systems where such processes are present~\cite{Porras_2013,Asenjo_2022,Asenjo_2023,Yelin_2025}.

\smallskip

\customsection{Conclusion \& Outlook}We have shown that states in Dicke superradiant decay can be written as positive mixtures of CSS  [Eq.~\eqref{eq:CSS_ansatz}]. This implies that there is no entanglement in the Dicke decay dynamics, i.e.,~the entanglement of formation is vanishing. The positive expansion coefficients $P_a$ ($a=0,\dots,N$) correspond to weights of CSS with polar angles $\theta_a$ and uniformly distributed azimuthal angles. Crucially, only the spacing between different $\theta_a$ needs to evolve time-dependently [single parameter, $\eta(t)$] to find positive $P_a > 0$. We then introduced an entanglement-optimized numerical QT unraveling algorithm, based on a simple random variation of a single parametrization angle in the Kraus representation of a discretized timestep. In the large $N$ limit we have found that this reproduces almost perfect CSS trajectories.

Identifying a closed analytical form for $\eta(t)$ for arbitrary $N$ would be particularly interesting. Indeed, the insight that TE can be transformed away paves a way to new efficient algorithms, also in cases  where, besides symmetric superradiant decay, Hamiltonian or dissipative processes without permutation symmetry are present. Here, combining $\phi$-randomization with matrix product states~\cite{schuch2008entropy, schollwock2011density} can lead to entanglement-optimized methods~\cite{vovk2022entanglement,vovk2024quantum,daraban2025non} for simulating collective dissipation on much larger scales than currently possible.

For future research, considering scenarios beyond the fully-symmetric Dicke limit has now become more tractable with our proposed method. Collective asymmetric jump operators can drive a separable quantum state into an entangled one in atomic systems with dipole-dipole interactions in free space or close to waveguides, for instance. These systems can generate multipartite entanglement, even without a coherent contribution of the master equation \cite{Joseph_Lee_2025}. For some particular configurations, this entanglement has been shown to be irrelevant, allowing, for example, the correct computation of two-point local observables in a large array of particles \cite{Xin_Zhang_2025}. With our method, such questions can now be explored on a full-state level: How much entanglement can be imprinted by a coherent contribution of the master equation or by asymmetric super- and sub-radiant decay channels?

\begin{acknowledgments} 
\customsection{Acknowledgments}We thank Claudiu Genes and Ruben Daraban for valuable discussions. We especially thank Nico Bassler for constructing the mathematical proof of separability in the related companion article~\cite{bassler2025absence}. This work was supported by the ERC Consolidator project MATHLOCCA (Grant nr.~101170485), by the CNRS through the EMERGENCE@INC2024 project DINOPARC, and by the French National Research Agency under the Investments of the Future Program project ANR-21-ESRE-0032 (aQCess). Computations  were  carried  out  using  resources  of  the High Performance Computing Center of the University of Strasbourg, funded by Equip@Meso (as part of the Investments for the Future Program) and CPER Alsacalcul/Big Data. P.R., A.C. and R.B. acknowledge the financial support of the São Paulo Research Foundation (FAPESP) (Grants No. 2022/12382-4, 2024/05564-4, 2022/06449-9, 2023/07463-8, 2023/14832-0, 2022/00209-6 and 2023/03300-7). 
L.O.R.S.~acknowledges the financial support of CAPES-COFECUB (CAPES, Grants Nos.~88887.711967/2022-00).
\end{acknowledgments}

\bibliography{main}

\pagebreak

\begin{widetext}

\newpage

\setcounter{equation}{0}
\setcounter{figure}{0}
\setcounter{table}{0}
\setcounter{page}{1}
\renewcommand{\theequation}{S\arabic{equation}}
\renewcommand{\thefigure}{S\arabic{figure}}

\appendix

\def\appendixname{}
\renewcommand{\thesection}{\Roman{section}}
\setcounter{secnumdepth}{1}

\makeatletter
\def\@seccntformat#1{\csname the#1\endcsname.\quad}

\renewcommand\section{\@startsection{section}{1}{\z@}%
  {-3.5ex \@plus -1ex \@minus -.2ex}%
  {2.3ex \@plus.2ex}%
  {\centering\normalfont\normalsize\bfseries}}
\makeatother

\begin{center}
{\large{ {\bf Supplemental Material for: \\ Unraveling Dicke Superradiant Decay with Separable Coherent Spin States}}}

\vskip0.5\baselineskip{P.~ Rosario$^{1,2}$, L.~O.~R.~Solak$^{1,2}$, A. Cidrim,$^{1}$, R. Bachelard$^{1}$, and J. Schachenmayer$^{2}$}

\vskip0.5\baselineskip{ {\it $^{1}$Departamento de F\'isica, Universidade Federal de S\~ao Carlos,\\ Rodovia Washington Lu\'is, km 235—SP-310, 13565-905 S\~ao Carlos, SP, Brazil\\
$^2$CESQ/ISIS (UMR 7006), CNRS and Universit\'{e} de Strasbourg, 67000 Strasbourg, France
}}

\end{center}

\section{Bipartite entanglement entropies in Dicke-state superpositions}
\label{app:dicke_states_entanglement}

Following \cite{latorre2005entanglement,schachenmayer2013entanglement} a Dicke state for $N$ emitters $\ket{N,m}$ (with $m$ denoting the number of excited states out of $N$ emitters) can be written as the tensor product of two smaller sub-spaces, we define block A and B with sizes $N-N_{B}$ and  $N_{B}$ respectively such that
\begin{align}
    \ket{N, m} = \sum_{l=0}^{N_{B}} \sqrt{p_{lm}} \ket{N_{B}, l}_A \ket{N-N_{B}, m-l}_B
    \label{eq:dicke_split}
\end{align}
the amplitudes are given by
\begin{align}
    p_{lm} = \frac{ {N_{B} \choose l} {N -N_{B} \choose m-l}  }{{N \choose m}}.
\end{align}
Here the states $\ket{N_{B}, k}_{A}$ and $\ket{N-N_{B}, k}_{B}$ are Dicke states of the emitters living in the ensembles A and B, respectively. Note that the sum in Eq.~\eqref{eq:dicke_split} includes invalid Dicke states, e.g.~when $m-l<0$ and $m-l > N-N_{B}$, however for those cases the $p_{lm}$ are zero. Taking now an arbitrary Dicke state superposition, we can write
\begin{align}
    \ket{\psi} 
    &= \sum_m c_{m} \ket{N, m} \\
    &= \sum_{m=0}^N c_m \sum_{l=0}^{N_{B}} \sqrt{p_{lm}} \ket{N_{B}, l}_A \ket{N-N_{B}, m-l}_B
\end{align}
now taking the partial trace with respect to to block A, we get
\begin{align}
   \hat \rho_B= \text{tr}_{A}\left(\ketbra{\psi}{\psi}\right)=  \sum_{k=0}^{N_{B}} \sum_{m=0}^N \sum_{n=0}^N c_m c^*_n  \sqrt{p_{kn}} \sqrt{p_{km}}\ket{N-N_{B}, m-k}_B
    \bra{N-N_{B}, n-k}_B
\end{align}
From a diagonalization of this matrix, the von Neumann entropy can be computed in the usual way. 

\section{Mapping matrix between Dicke and CSS representation}
\label{app:details_on_mapping}

Spin coherent states are characterized by a collection of individual particles pointing in the same direction, no entanglement is present. We formally write
\begin{align}
    \ket{\theta(t),\phi}=\bigotimes_{j=1}^{N}\left[\cos\left(\frac{\theta(t)}{2}\right)\ket{e}+\sin\left(\frac{\theta(t)}{2}\right)\me^{\mi\phi}\ket{g}\right]_{j}
\end{align}
with $0\leq \theta(t) \leq \pi$ and $0\leq \phi \leq 2\pi$. In terms of Dicke states $\{\ket{m}\}_{m=0}^{N}$ with $m$ being the number of excitations, $\ket{\theta(t),\phi}$ can also be written as
\begin{align}
    \ket{\theta(t),\phi}=\sum_{k=0}^{N}\sqrt{\begin{pmatrix}
 N\\
k  
\end{pmatrix}}\left(\cos\left(\frac{\theta(t)}{2}\right)\right)^{N-k}\left(\sin\left(\frac{\theta(t)}{2}\right)\right)^{k}\me^{\mi k\phi}\ket{N-k}
\label{eq:coherent_state}
\end{align}
those states live on the surface of the generalized Bloch sphere. On the other hand, a $N$-qubit statistical mixture of Dicke states, is defined as
\begin{align}
    \hat{\rho}(t)=\sum_{m=0}^{N}\rho_{m}(t)\ketbra{m}{m}
    \label{eq:app_mix_dicke}
\end{align}
where for  some specific populations distributions $\rho_{m}(t)$, the quantum state in Eq.~\eqref{eq:app_mix_dicke} can be full separable \cite{Nengkun_separability_2016}. In the following, we show that a statistical mixture of Dicke states similar to Eq.~\eqref{eq:app_mix_dicke} can be obtained by combining multiple spin coherent states. For that, we propose the time-dependent ansatz
\begin{align}
    \hat{\rho}(t)&=\sum_{a=0}^{N_{\theta}}P_{a}(t)\left[\frac{1}{N_{\phi}}\sum_{b=1}^{N_{\phi}}\ketbra{\theta_{a}(t),\phi_{b}}{\theta_{a}(t),\phi_{b}}\right]
    \label{eq:app_coh_mix}
\end{align}
with $N_{\theta}$ and $N_{\phi}$ defining how many angles $\theta$ and $\phi$ are needed to reproduce the target state in Eq.\eqref{eq:app_mix_dicke}. Therefore, choosing $N_{\phi}=2N$, $N_{\theta}=N$ and $\phi_{b}=\frac{2\pi}{N_{\phi}}b=\frac{\pi}{N}b$, we arrive to the statistical mixture of Dicke states
\begin{align}
  \nonumber  \hat{\rho}& =\sum_{m=0}^{N}\left[\sum_{a=0}^{N_{\theta}}P_{a}(t)\begin{pmatrix}
 N\\
m  
\end{pmatrix}\left(\cos\left(\frac{\theta_{a}(t)}{2}\right)\right)^{2N-2m}\left(\sin\left(\frac{\theta_{a}(t)}{2}\right)\right)^{2m}\right]\ketbra{N-m}{N-m},\\
&=\sum_{m=0}^{N}\left[\sum_{a=0}^{N}P_{a}(t)\begin{pmatrix}
 N\\
m  
\end{pmatrix}z_{a}(t)^{N-m}\left(1-z_{a}(t)\right)^{m}\right]\ketbra{N-m}{N-m} \ \ \text{with}\ \ z_{a}(t)=\cos^{2}\left(\frac{\theta_{a}(t)}{2}\right).
\label{eq:app_cohe_theta}
\end{align}
note that, the quantity inside the brackets can be written as the product of a stochastic matrix times a probability vector as $\bm M \cdot \bm P$, with
\begin{align}
    &\bm P=[P_{0}(t),P_{1}(t),\hdots,P_{N}(t)]^{T} \ \ (\bm{M})_{mn} =\begin{pmatrix}
 N\\
m  
\end{pmatrix}z_{n}(t)^{N-m}[1-z_{n}(t)]^{m}, \ \ m,n=0,1,\hdots,N.
\end{align}
Therefore,  any probability distribution obtained from the product $\bm M \cdot \bm P$ corresponds to a full separable state iff $\bm P \geq 0$.

\section{Exact expression of $\eta(t)$ for $N=2$}
\label{app:arccos}
Defining the general coherent state representation
\begin{align}
    \bm{P}^{[\tau]} = \left(\bm{M}^{[\tau]}\right)^{-1} \cdot  \bm{\rho}^{[\tau]},
\end{align}
for a given $\bm{\rho}^{[\tau]}$, we need to find $\eta(t)$ such that $\forall \tau$ $\bm{P}^{[\tau]}\geq 0$. For the case $N=2$, taking $\theta_{0}=0$, $\theta_{1}=\eta(t)\pi/2$ and $\theta_{2}=\eta(t)\pi$ we get the set of inequalities (here, $\Gamma \equiv 1$)
\begin{align}
    &P_{1}=\frac{1}{8}\left(8\me^{t}+(8-8\me^{t}+3t)\csc^{2}\left(\frac{\eta(t)\pi}{4}\right)+2(-1+\me^{t}-t)\csc^{4}\left(\frac{\eta(t)\pi}{4}\right)-t\sec^{2}\left(\frac{\eta(t)\pi}{4}\right)\right)\geq 0\\
    &P_{2}=-\frac{\me^{-t}\left(-1+\me^{t}-\frac{3}{2}t+(-1+\me^{t}-\frac{t}{2})\cos(\eta(t)\pi)\right)\csc^{4}\left(\frac{\eta(t)\pi}{4}\right)}{2\left(1+2\cos\left(\frac{\eta(t)\pi}{2}\right)\right)}\geq 0\\
    &P_{3}=\frac{\me^{-t}\csc^{4}\left(\frac{\eta(t)\pi}{4}\right)\left(-2+2\me^{t}-t-t\sec^{2}\left(\frac{\eta(t)\pi}{4}\right)\right)}{8\left(1+2\cos\left(\frac{\eta(t)\pi}{2}\right)\right)}\geq 0
\end{align}
taking the mapping $y=\cos^{2}(\eta(t)\pi/2)$ and using Chebyshev polynomials properties, we arrive to the expression presented in the main text for $\eta(t)$. When $N>2$, the number of inequalities increases like $N+1$ and obtaining $\eta(t)$ becomes analytically unfeasible. However, this case is numerically addressed in the main text.

\section{Numerical considerations and additional data}
\label{app:numerics}

\begin{figure*}
    \centering
    \includegraphics[width=\textwidth]{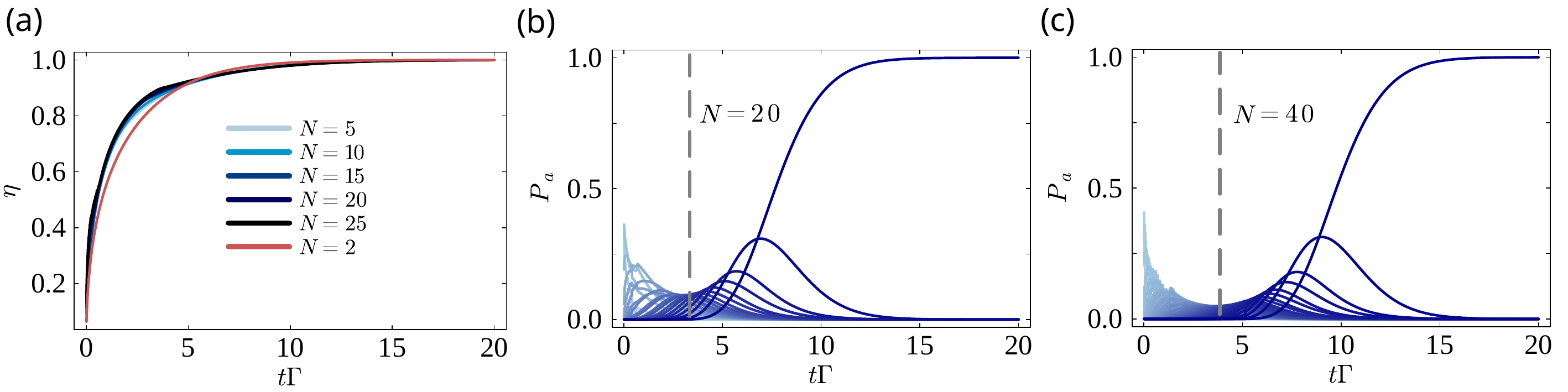}
    \caption{(a) Numerically extracted positive-$\bm{P}$ solutions close to the lower passage for different values of $N$. All solutions remain close to the analytical result for $N=2$ (red line). The smooth behavior may suggest the existence of a a general smooth solution for $N>2$. (b) $P_a$ evolution for $N=20$ and (c) $N=40$ ($\mathcal{N} < 10^{-6}$). The vertical dashed lines indicate the burst time. \label{fig:extra_data}}
\end{figure*}

\customsection{Lower passage $\eta(t)$ solutions} At each time, $\bm{P}$ can be obtained by solving the linear system Eq.~\eqref{eq:linsys} while varying the parameter $\eta(t)$. We find that in this way negativites along the evolution can be made arbitrary small up to numerical precision limits. We search for solutions of $\eta(t)$ following the lower passage. For the plots in this work we show corresponding evolutions of $\bm{P}$ using values with  negativities $\mathcal{N} < 10^{-6}$. Note that in all cases for numerical stability we truncate probabilities in the vector $\rho_t$ below relative machine precision ($10^{-16}$). We show some selected evolution of $\eta(t)$ extracted numerically in Fig.~\ref{fig:extra_data}(a). It is important to point out that there is only a little variation in the solutions for different $N$ and that the solutions remains rather close to the exact analytical curve for $N=2$. In Fig.~\ref{fig:extra_data}(b/c) we also show the evolution of $\bm{P}$ for $N=20$ and $N=40$, respectively, together with the burst time $t_B$ as vertical dashed line. Note that not in all cases we found smooth $\eta(t)$ and $P_a(t)$, since multiple solutions exist and our numerical optimizer can jump between them.

\bigskip

\customsection{Independence of TE on the choice $\theta_F$ and $\Delta t$} For the simulations related to transformed Kraus operators according to~Eq.~\eqref{eq:kraus_F}, in Fig.~\ref{fig:TE_verify}(a) we verify that selecting different values of $\theta_{F}$ have no major impact on the TE evolution. In addition we verify in Fig.\ref{fig:TE_verify}(b) that our QT simulations are converged in the timestep $\Delta t$ and do not depend on $\Delta t$.

\begin{figure*}
    \centering
    \includegraphics[width=0.7\textwidth]{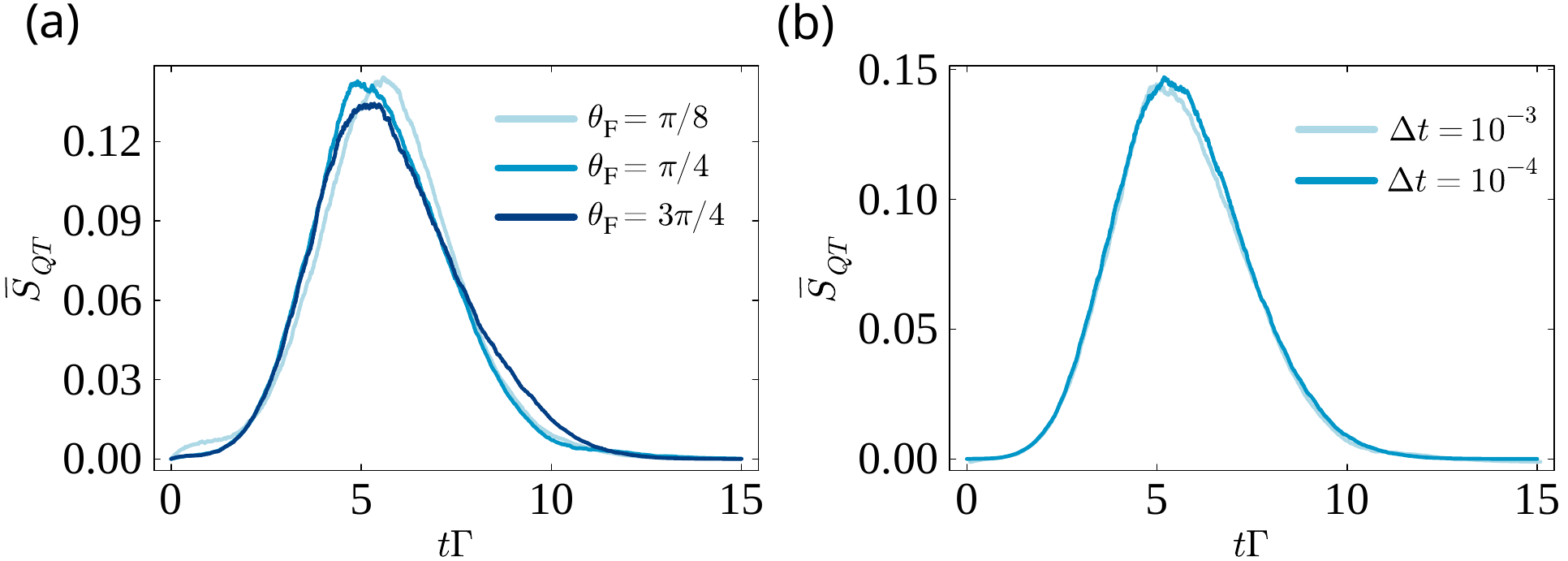}
    \caption{QT simulations for $N=50$ with randomized $\phi_F$  taking the average over 100 trajectories: (a) TE evolution using various choices of $\theta_F$. No major variation in TE is observed when comparing $\theta_F=\pi/8, \pi/4, 3\pi/4$; (b) TE evolution for different timesteps, showing invariance with respect to $\Delta t$\label{fig:TE_verify}}
\end{figure*}

\end{widetext}
\end{document}